\documentclass[12pt]{iopart}
\usepackage{amssymb,graphicx}
\usepackage{times} 
\begin{document}
%%%%%%%%%%%FIGURES%%%%%%%%%%%%%%%%%%
%
\newcommand{\fig}[2]{\includegraphics[width=#1]{./figures/#2}}
\newcommand{\Fig}[1]{\includegraphics[width=\columnwidth]{./figures/#1}}
\newlength{\bilderlength}
\newcommand{\bilderscale}{0.35}
\newcommand{\storebilderscale}{\bilderscale}
\newcommand{\bilderskip}{\hspace*{0.8ex}}
\newcommand{\textdiagram}[1]{%
\renewcommand{\bilderscale}{0.25}%
\diagram{#1}\renewcommand{\bilderscale}{\storebilderscale}}
\newcommand{\diagram}[1]{%
\settowidth{\bilderlength}{\bilderskip%
\includegraphics[scale=\bilderscale]{./figures/#1}\bilderskip}%
\parbox{\bilderlength}{\bilderskip%
\includegraphics[scale=\bilderscale]{./figures/#1}\bilderskip}}
\newcommand{\Diagram}[1]{%
\settowidth{\bilderlength}{%
\includegraphics[scale=\bilderscale]{./figures/#1}}%
\parbox{\bilderlength}{%
\includegraphics[scale=\bilderscale]{./figures/#1}}}
%
%%%%%%%%%% Some new commands %%%%%%%%%%%
%\newcommand{\rme}{{\mathrm{e}}}
%\newcommand{\rmd}{{\mathrm{d}}} 
\newcommand{\half}{\frac12}
%%%%%%%%%%%%%%%%%%%%%%%%%%%%%%%%%%%%%%%%%%%%%%%%%%%%%%%%%%%%%%%%%%%%%%%%
%Global variables
%
%\bibliographystyle{../../macros/revtex4/apsrmp}
%\bibliographystyle{../macros/KAY}
%\bibliographystyle{../../macros/KAY.bst}
%\bibliographystyle{unsrt}
%%%%%%%%%%%%%%%%%%%%%%%%%%%%%%%%%%%%%%%%%%%%%%%%%%%%%%%%%%%%%%%%%%%%%%%%%%%%
%\begin{frontmatter} 
%

\title[Susy Breaking in Disordered Systems and Relation to FRG and RSB]{Supersymmetry Breaking in Disordered
Systems  and  Relation to  Functional
Renormalization and  Replica-Symmetry Breaking}

\author{Kay J\"org Wiese}
\address{Laboratoire de Physique Th\'eorique de l'Ecole Normale 
Sup\'erieure, 24 rue Lhomond, 75005 Paris, France}

\begin{abstract}
In this article, we study an elastic manifold in quenched disorder in
the limit of zero temperature. Naively it is equivalent to a free
theory with elasticity in Fourier-space proportional to $ k^{4}$
instead of $k^{2}$, i.e.\ a model without disorder in two
space-dimensions less. This phenomenon, called dimensional reduction,
is most elegantly obtained using supersymmetry. However, scaling
arguments suggest, and functional renormalization shows that
dimensional reduction breaks down beyond the Larkin length. Thus one
equivalently expects a break-down of supersymmetry. Using methods of
functional renormalization, we show how supersymmetry is broken. We
also discuss the relation to replica-symmetry breaking, and how our
formulation can be put into work to lift apparent ambiguities in
standard functional renormalization group calculations.

\noindent 
Dedicated to Lothar Sch\"afer at the occasion of
his 60th birthday.
\end{abstract}

%contents

%%%%%%%%%%%%%%%%%%%%%%%%%%%%%%%%%%%%%%%%%%%%%%%%%%%%%%%%%%%%%%%%%%
%%%%%%%%%%%%%%%%%%%%%%%%%%%%%%%%%%%%%%%%%%%%%%%%%%%%%%%%%%%%%%%%%%
%%%%%%%%%%%%%%%%%%%%%%%%%%%%%%%%%%%%%%%%%%%%%%%%%%%%%%%%%%%%%%%%%%
%%%%%%%%%%%%%%%%%%%%%%%%%%%%%%%%%%%%%%%%%%%%%%%%%%%%%%%%%%%%%%%%%%
%%%%%%%%%%%%%%%%%%%%%%%%%%%%%%%%%%%%%%%%%%%%%%%%%%%%%%%%%%%%%%%%%%
%%%%%%%%%%%%%%%%%%%%%%%%%%%%%%%%%%%%%%%%%%%%%%%%%%%%%%%%%%%%%%%%%%
%%%%%%%%%%%%%%%%%%%%%%%%%%%%%%%%%%%%%%%%%%%%%%%%%%%%%%%%%%%%%%%%%%
%%%%%%%%%%%%%%%%%%%%%%%%%%%%%%%%%%%%%%%%%%%%%%%%%%%%%%%%%%%%%%%%%%
%%%%%%%%%%%%%%%%%%%%%%%%%%%%%%%%%%%%%%%%%%%%%%%%%%%%%%%%%%%%%%%%%%
%%%%%%%%%%%%%%%%%%%%%%%%%%%%%%%%%%%%%%%%%%%%%%%%%%%%%%%%%%%%%%%%%%
%                                                                %
%                        Introduction                            %
%                                                                %
%%%%%%%%%%%%%%%%%%%%%%%%%%%%%%%%%%%%%%%%%%%%%%%%%%%%%%%%%%%%%%%%%%
%%%%%%%%%%%%%%%%%%%%%%%%%%%%%%%%%%%%%%%%%%%%%%%%%%%%%%%%%%%%%%%%%%
%%%%%%%%%%%%%%%%%%%%%%%%%%%%%%%%%%%%%%%%%%%%%%%%%%%%%%%%%%%%%%%%%%
%%%%%%%%%%%%%%%%%%%%%%%%%%%%%%%%%%%%%%%%%%%%%%%%%%%%%%%%%%%%%%%%%%
%%%%%%%%%%%%%%%%%%%%%%%%%%%%%%%%%%%%%%%%%%%%%%%%%%%%%%%%%%%%%%%%%%
%%%%%%%%%%%%%%%%%%%%%%%%%%%%%%%%%%%%%%%%%%%%%%%%%%%%%%%%%%%%%%%%%%
%%%%%%%%%%%%%%%%%%%%%%%%%%%%%%%%%%%%%%%%%%%%%%%%%%%%%%%%%%%%%%%%%%
%%%%%%%%%%%%%%%%%%%%%%%%%%%%%%%%%%%%%%%%%%%%%%%%%%%%%%%%%%%%%%%%%%
%%%%%%%%%%%%%%%%%%%%%%%%%%%%%%%%%%%%%%%%%%%%%%%%%%%%%%%%%%%%%%%%%%
\section{Introduction} The statistical mechanics of even
well-understood physical systems subjected to quenched disorder still
poses major challenges. For a large class of these systems, as e.g.\
random-field models or elastic manifolds in quenched disorder, an
apparent simplification appears: Supposing that all moments of the
disorder are finite, one can show that all correlation functions in
the disordered model, in the limit of zero temperature, are equivalent
to those of the pure system at finite temperature in two
space-dimensions less, at a temperature proportional to the second
moment of the quenched disorder. This phenomenon is called dimensional
reduction (DR) \cite{EfetovLarkin1977}. The most elegant way to prove
it is to use the supersymmetry approach \cite{ParisiSourlas1979}, as
we will detail below. However, one also knows that dimensional
reduction gives the wrong result at large scales, more precisely at
scales larger than the Larkin length. The latter is obtained from an
Imry-Ma type argument due to Larkin, balancing elastic energy and
disorder, as we detail below. For a $d$-dimensional elastic manifold
in quenched disorder, the elastic and disorder energy are
\begin{equation}\label{Es} E_{\mathrm{el}}[u]=\int \rmd^{d} x\,
\half (\nabla u(x))^{2} \ , \qquad E_{{\mathrm{DO}}}[u]=\int \rmd^{d}
x\, V(x,u(x))\ .
\end{equation}
For $d=1$, these are polymers, for which a lot is known
\cite{SchaeferBuch}; for $d=2$ membranes; and for $d=3$ elastic
crystals, as e.g.\ charge-density waves.  For simplicity we consider
disorder which at the microscopic scale is Gaussian and short-ranged
with second moment
\begin{equation}\label{shortrangeDO}
\overline{V(x,u) V(x',u')} = \delta^{d}(x-x') R(u-u')\ .
\end{equation}
Long-range correlated disorder $R(u)$ is possible, and leads in
general to a different universality class. This will play no role in
the following.  The most important observable is the roughness
exponent $\zeta$, which describes the scaling of the 2-point function
\begin{equation}\label{zetadef}
\overline{\left[u(x)-u(x') \right]^{2}} \sim |x-x'|^{2\zeta}\ .
\end{equation}
The Larkin argument compares, as a function of system size $L$,
elastic energy $E_{\mathrm{el}} \sim L^{d-2}$ and disorder energy
$E_{\mathrm{DO}} \sim L^{d/2}$ to conclude that in dimensions smaller
than four, disorder always wins at large scales, leading to an RG-flow
to strong coupling (in a way to be made more precise below). This
suggests that the dimensional reduction result, derived below via the
Supersymmetry method, 
\begin{equation}\label{DR}
\overline{u_{k}u_{-k}}=\frac{-R'' (0)}{(k^{2})^{2}}\qquad
\Longrightarrow \qquad \zeta_{\mathrm{DR}} = \frac{4-d}2
\end{equation}
will become  incorrect below four dimensions. 
 
\section{The functional RG treatment} In this section we review some
important points of the functional RG treatment, which will facilitate
the derivation of the corresponding formulas in the supersymmetric
treatment. Functional RG was first introduced in
\cite{WilsonKogut1974,WegnerHoughton1973}, and pioneered for the
problem at hand in
\cite{Fisher1985b,DSFisher1986,NattermanStepanowTangLeschhorn1992,%
NarayanDSFisher1992a},
to  cite the earliest contributions. Important improvements \cite{NarayanDSFisher1993a,BalentsDSFisher1993,LeschhornNattermannStepanowTang1997,BucheliWagnerGeshkenbeinLarkinBlatter1998,ChauveLeDoussalWiese2000a,ScheidlDincer2000,Scheidl2loopPrivate,ChauveLeDoussal2001,LeDoussalWiese2001,LeDoussalWieseChauve2002,LeDoussalWiese2002a,GorokhovFisherBlatter2002,GlatzNattermannPokrovsky2002,FedorenkoStepanow2002,LeDoussalWiese2003a,RossoKrauthLeDoussalVannimenusWiese2003,LeDoussalWieseChauve2003,LeDoussalWiese2003b,BalentsLeDoussal2003,LeDoussalWiese2004a,BalentsLeDoussal2004} have
been obtained by several authors, see \cite{Wiese2003a} for a more
detailed introduction and review.

The Larkin argument suggests that four is the upper critical
dimension and that an $\epsilon$-expansion \cite{SchaeferBuch} with
\begin{equation}
\label{epsilon}
\epsilon=4-d
\end{equation}
about dimension four is possible. Taking the dimensional reduction
result (\ref{DR}) in $d=4$ dimensions tells us that the field $u$ is
dimensionless. Thus, the width $\sigma = -R''(0)$ of the disorder is
not the only relevant coupling at small $\epsilon$, but any function
of $u$ has the same scaling dimension in the limit of $\epsilon=0$,
and might thus equivalently contribute. The natural consequence is
that one has to follow the full function $R(u)$ under renormalization,
instead of just its second moment $R''(0)$. Such an RG-treatment is
most easily implemented in the replica approach: The $n$ times
replicated partition function becomes after averaging over disorder
\begin{eqnarray}\label{reps}
&&\overline{ \exp\!\left(-\frac{1}T \sum_{a=1}^{n}
E_{\mathrm{el}}[u_{a}] - \frac{1}T \sum_{a=1}^{n} {E_{\mathrm{DO}}
[u_{a}]} \right) } \nonumber\\&&
\qquad = \exp \left(-\frac1T \sum_{a=1}^{n} {
E_{\mathrm{el}}[u_{a}]} + \frac1{2T^{2}} \sum_{a,b=1}^{n} \int \rmd^d
x\, R\Big(u_{a}(x)-u_{b}(x)\Big) \right) \ .
\end{eqnarray}
Perturbation theory is constructed along the following lines (see
\cite{BalentsDSFisher1993,LeDoussalWieseChauve2003} for more details.)  The
bare correlation function, graphically depicted as a solid line, is
with momentum $k$ flowing through and replicas $a$ and $b$
\begin{equation}
_{a}\diagram{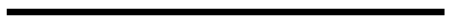}_{b}\ = \frac {T\delta_{ab}} {k^{2}}\ .
\end{equation}
The disorder vertex is 
\begin{equation}
\stackrel{\!\!\!x}{\diagram{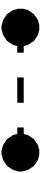}_{b}^{a}}\ =\int_{x}\sum_{a,b}
R\Big(u_{a}(x)-u_{b}(x)\Big)\ .
\end{equation}
The rules of the game are to find all contributions which correct $R$,
and which survive in the limit of $T\to 0$. At leading order, i.e.\ order
$R^{2}$, counting of factors of $T$ shows that only the terms with one
or two correlators contribute. On the other hand, $\sum_{a,b}
R(u_{a}-u_{b})$ has two independent sums over replicas. Thus at order
$R^{2}$ four independent sums over replicas appear, and in order to
reduce them to two, one needs at least two correlators (each
contributing a $\delta_{ab}$). Thus, at leading order, only diagrams
with two propagators survive. These are the following (noting $C(x-y)$
the Fourier transform of $1/k^{2}$):
\begin{eqnarray}\fl
\parbox{0mm}{\rule{0mm}{5.3mm}}^{a}_{b}\!\!\stackrel{\!\!\!x\hspace{1.4cm}y}{\diagram{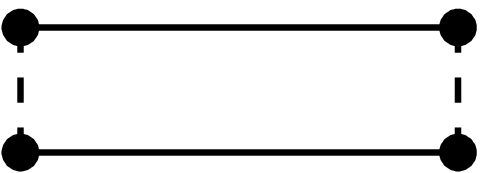}_{b}^{a}} &=& \int_{x}
 R''(u_{a}(x) -u_{b}(x))R''(u_{a}(y) -u_{b}(y))  C(x-y)^{2}
\\ 
\fl \parbox{0mm}{\rule{0mm}{5.3mm}}^{a}_{a}\!\!\stackrel{\!\!\!x\hspace{1.4cm}y}{\diagram{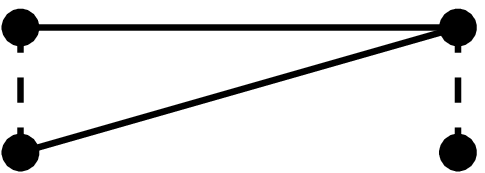}_{b}^{a}}
&=& - \int_{x} R''(u_{a}(x) -u_{a}(x))R''(u_{a}(y) -u_{b}(y))
C(x-y)^{2} \ .
\end{eqnarray}
In a renormalization program, we are looking for the divergences of
these diagrams. These divergences are localized at $x=y$, which allows
to approximate $R''(u_{a}(y)-u_{b}(y))$ by $R''(u_{a}(x)-u_{b}(x))$.
The integral $\int_{x-y} C(x-y)^{2} = \int_{k} \frac 1 {( k^{2}+m^{2})^{2}} =
\frac {m^{-\epsilon}}\epsilon $ (using the most convenient
normalization for $\int_{k}$), is the standard
1-loop diagram, which we have chosen to regulate in the infrared by a
mass, i.e.\ physically by a harmonic well which is seen by the
manifold.

Note that the following diagram also contains two correlators (correct
counting in powers of temperature), but is not a 2-replica but a
3-replica sum:
\begin{equation}
{\raisebox{-1mm}{$\parbox{0mm}{\rule{0mm}{5.3mm}}^{a}_{b}$}}\!\!\stackrel{\!\!\!x\hspace{1.4cm}y}{\diagram{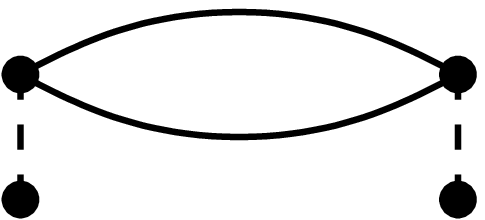}}\!\!
{\raisebox{-1mm}{$\parbox{0mm}{\rule{0mm}{5.3mm}}^{a}_{c}$}}\ \ .
\end{equation}

Taking into account the combinatorial factors, and a rescaling of $R$
(which remember has dimension $\epsilon$ for a dimensionless field
$u$) as well as of the field $u$ (its dimension being the roughness
exponent $\zeta$), we arrive at
\begin{equation}\label{RG1loop}
-m\frac{\partial}{\partial m} R(u) = (\epsilon -4 \zeta) R(u) + \zeta
u R'(u) + \half R''(u)^{2} - R''(u) R''(0)\ .
\end{equation} 
\begin{figure}
\centerline{\fig{12cm}{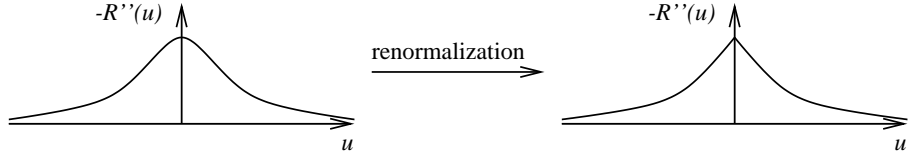}}
\caption{Change of $-R'' (u)$ under renormalization and formation of
the cusp.} \label{fig:cusp}
\end{figure}%
Note that the elasticity does not get renormalized due to the statistical
tilt symmetry $u (x)\to  u (x)+ \alpha x$.

The crucial observation is that when starting with
smooth microscopic disorder, integration of the RG-equation leads to a
cusp in the second derivative of the renormalized disorder at the
Larkin-length, as depicted on figure \ref{fig:cusp}.  This can easily
be seen from the flow-equation of the fourth derivative (supposing
analyticity), which from (\ref{RG1loop}) is obtained as
\begin{equation}\label{R4(0)flow}
-m\frac{\partial}{\partial m} R''''(0) = \epsilon R''''(0) + 3
R''''(0)^{2} \ .
\end{equation}
(Note that this explains also the appearance of the combination
$\epsilon-4\zeta$ in (\ref{RG1loop})). This equation has a singularity
$R'''' (0)=\infty $ after a finite renormalization time, equivalent to
the appearance of the cusp, as depicted on figure
\ref{fig:cusp}. After that dimensional reduction (\ref{DR}) is no
longer valid. This can most easily be seen from the flow of $R''
(0)$: Deriving (\ref{RG1loop}) twice w.r.t.\ $u$, and then taking the
limit of $u\to 0$ leads to 
\begin{equation}\label{flowR''0}
-\frac{\partial}{\partial m} R'' (0) = (\epsilon -2\zeta) R'' (0) +
R''' (0^{+})^{2}\ .
\end{equation}
In the analytic regime $R''' (0^{+})=0$, such that the fixed-point
condition $-\frac{\partial}{\partial m} R'' (0)=0$ implies $\zeta
=\frac{\epsilon}{2}\equiv \frac{4-d}{2}$; after appearance of the
cusp, $R''' (0^{+})\neq 0$, thus  $\zeta$ has to change. 

This analysis can be continued to higher orders.  Let us cite some
key results at 2-loop
order\cite{ChauveLeDoussalWiese2000a,LeDoussalWieseChauve2003}, for
which the RG-equation reads
\begin{eqnarray}\label{2loopRG}
- m\frac{\partial}{\partial m} R (u) &=& \left(\epsilon -4 \zeta
\right) R (u) +
\zeta u R' (u) + \frac{1}{2} R'' (u)^{2}-R'' (u)R'' (0) \nonumber \\
&& + \frac{1}{2}\left(R'' (u)-R'' (0) \right)R'''
(u)^{2}-\frac{1}{2}R''' (0^{+})^{2 } R'' (u) \ .
\end{eqnarray}
Different microscopic disorder leads to different RG fixed points. The
latter are solutions of equation (\ref{2loopRG}), with $- m
\frac\partial{\partial m} R(u)=0$; it is important to note that given
a microscopic disorder, the exponent $\zeta$, solution of
(\ref{2loopRG}) is unique.  For random-bond disorder (short-ranged
potential-potential correlation function) the result is $\zeta = 0.208
298 04 \epsilon +0.006858 \epsilon ^{2}$.  In the case of random field
disorder (short-ranged force-force correlations) $\zeta
=\frac{\epsilon }{3}$. Both results compare well with numerical
simulations.

One should also note that (\ref{2loopRG}) contains a rather peculiar
``anomalous term'', namely $R'''(0^{+})^{2}$, which only appears after
the occurrence of the cusp. This term is in general hard to get, since
the calculation naturally gives factors of $R'''(0)$, which are 0 by
parity, and not $R'''(0^{+}) = -R'''(0^{-})$. Several procedures to
overcome these apparent ambiguities have been developed
\cite{LeDoussalWieseChauve2003}. Supersymmetry will allow for another
prescription, as will be discussed below.

\section{Supersymmetry and its Breaking}\label{a:susy} Another way to
average over disorder is to use additional fermionic degrees of
freedom. %which
%deliver additional factors of roughly the inverse partition function,
It is more commonly referred to as the supersymmetric
method. Supersymmetry is manifest using one copy of the system, where
it immediately leads to dimensional reduction, as we show
below. However it can not account for the non-trivial physics due to
the appearance of the cusp and the corresponding breakdown of
dimensional reduction, and supersymmetry. This is possible when
considering $n\neq 1$ copies of the system, with $n=2$ being
completely legitimate. Here we give a general formulation, in which
one can either discard the fermionic degrees of freedom, thus
reconstructing a replica formulation at $n=0$, or set $n$ to e.g.\
$n=1$, thus exploring supersymmetry.

Define
\begin{equation}
{\cal H}[u_{a},j_{a},V] = \sum_{a=1}^{n}\int_{x}\half \left(\nabla u_{a} (x)
\right)^{2} + V (x,u_{a} (x)) + j_{a} (x)u_{a} (x)
\ .
\end{equation}
Then the normalized generating function of correlation functions for a
given disorder $V$ is
\begin{equation}\label{su2}
{\cal Z} [ j] := \frac{\int \prod_{a}{\cal D}[{u_{a}}
]\rme^{-\frac{1}{T}{\cal H} [{u_{a}}, {j_{a}},{V}]}}{\int \prod_{a} {\cal
D}[{u_{a} } ]\rme^{-\frac{1}{T}{\cal H} [{u_{a} },{0},{V} ] }}
\ .
\end{equation}
In the limit of $T\to 0$ only configurations which minimize the energy
survive; these configurations satisfy $\frac{\delta{\cal H}[{u_{a} },
{j_{a}},{V} ] }{\delta {u_{a} } (x) }=0$, of which we want to insert a
$\delta$-distribution in the path-integral. This has to be accompanied
by a factor of $\det\! \left[\frac{\delta^{2}{\cal H}[{u_{a} },
{j_{a}},V ] }{\delta {u_{a} } (x)\delta {u_{a} } (y) }\right]$, such
that the integral over this configuration is normalized to 1, and
supposing only a single configuration, the denominator can be dropped,
leading to
\begin{equation}\label{su4}
{\cal Z} [ j] = \int\prod_{a} {\cal D}[{u_{a}}]\ \delta\! \left[
\frac{\delta{\cal H}[{u_{a} }, {j_{a}},{V} ] }{\delta {u_{a} } (x) }
\right]\, \det\! \left[\frac{\delta^{2}{\cal H}[{u_{a} }, 0 ,V ]
}{\delta {u_{a} } (x)\delta {u_{a} } (y) } \right] \ .
\end{equation}
Note that we neglect problems due to multiple minima, maxima, or
saddle points. These configurations are incorrectly contained in
(\ref{su4}), and are usually blamed for the failure of the
supersymmetry approach. We will comment on this point later.  For the
moment, we continue with (\ref{su4}) and see how far we can get. 
Using an imaginary auxiliary field $\tilde u (x)$ and two
anticommuting Grassmann fields $\bar \psi (x)$ and $\psi (x)$ (per
replica), this can be written as
\begin{eqnarray}\label{su5}
\fl {\cal Z} [ j] & =& \int\prod_{a} {\cal D}[{u_{a} }] {\cal D}[{\tilde
u_{a} }] {\cal D}[{\bar \psi_{a} }] {\cal D}[{\psi}_{a}]\,\nonumber \\
\fl &&  \qquad \exp\!
\left[-{ \int_{x}\tilde u_{a} (x)\frac{\delta{\cal H}[{u_{a} },
{j_{a}},V ]} {\delta {u_{a} } (x) }+\bar \psi_{a}(x)\frac{\delta^{2}
{\cal H}[{u_{a} }, {j_{a}},{V} ] }{\delta {u_{a} } (x)\delta {u_{a} }
(y) } \psi_{a} (y) } \right] \ .
\end{eqnarray}
Averaging over disorder yields with the force-force correlator $\Delta
(u):=-R'' (u)$
\begin{eqnarray}\label{su6a}
 \overline{{\cal Z} [ j]}&=&\int\prod_{a} {\cal D}[{u_{a} }] {\cal
D}[{\tilde u_{a} }] {\cal D}[{\bar \psi_{a} }] {\cal D}[{\psi}_{a}]
\exp \left(-{\cal S}[u_a,\tilde u_{a},\bar \psi_{a}, \psi_{a}, j_{a}]
\right)
  \nonumber \\
\fl\label{su6b} {\cal S}[u_a,\tilde u_{a},\bar \psi_{a}, \psi_{a},j_{a}]
&=& \sum_{a} \int_{x} \tilde u_{a} (x) (-\nabla^{2} u_{a} (x) +j_{a}
(x)) + \bar \psi_{a} (x)
(-\nabla^{2})\psi_{a} (x) \nonumber \\
&&- \sum_{a,b} \int_{x} \Big[ \half \tilde u_{a} (x)\Delta (u_{a} (x)-u_{b}
(x))\tilde u_{b} (x)\nonumber \\
&& \qquad\quad+ \half \bar \psi_{a} (x)\psi _{a} (x)\Delta''
(u_{a} (x)-u_{b} (x))\bar \psi_{b} (x)\psi_{b} (x) \nonumber \\
&&\qquad\quad - \tilde u_{a} (x)
\Delta' (u_{a} (x)-u_{b} (x)) \bar \psi_{b} (x)\psi_{b} (x)\Big]
\ .
\end{eqnarray}
We first analyze $n=1$. Suppose that $\Delta (u)$ is even and analytic
to start with, then only the following terms survive from (\ref{su6b})
\begin{equation}\label{reallySusy}\fl
{\cal S}_{\mathrm{Susy}}[u,\tilde u,\bar\psi,\psi,j]= \int_{x} \tilde u
(x) (-\nabla^{2 }u (x) +j (x)) + \bar \psi (x)
(-\nabla^{2})\psi (x) - \half \tilde u (x)\Delta (0)\tilde
u (x)
\end{equation}
(We have used that $\bar \psi_{a}^{2}=\psi_{a}^{2}=0$ to get rid of the
4-fermion-term.) This action possesses a supersymmetry, which
is most manifest when grouping terms together into a superfield 
\begin{equation}\label{superfielddef}
U (x,\bar \Theta ,\Theta) = u (x)+ \bar \Theta \psi (x)+\bar \psi (x)
\Theta + \Theta \bar \Theta \tilde u (x)
\ .
\end{equation}
The action (\ref{reallySusy}) can then be written with the
SuperLaplacian $\Delta_{s}$ as 
\begin{equation}\fl
{\cal S}_{\mathrm{Susy}}= \int \rmd \Theta \rmd \bar \Theta\int_{x} U
(x,\bar \Theta ,\Theta) (\Delta_{s}) U (x,\bar \Theta ,\Theta)\ ,
\qquad \Delta_{s} := \nabla^{2}-\Delta (0) \frac{\partial}{\partial
\bar \Theta}\frac{\partial}{\partial \Theta}
\end{equation}
and is invariant under the action of the supergenerators 
\begin{equation}
Q := x \frac{\partial}{\partial \Theta}-\frac{2}{\Delta (0)} \bar
\Theta \nabla \ , \qquad \bar Q:=x \frac{\partial}{\partial \bar
\Theta}+\frac{2}{\Delta (0)} \Theta \nabla\ .
\end{equation}
Since ``bosons'' $u$ and $\tilde u$, and ``fermions'' $\bar \psi$ and
$\psi$ only appear to quadratic order, all expectation values are
trivially Gaussian. Especially is 
\begin{equation}\label{ukumk}
\overline{u_{k} u_{-k}} = \frac{\Delta (0)}{(k^{2})^{2}} \ , 
\end{equation}
which is the result cited in (\ref{DR}), recalling that $\Delta (u)=
-R'' (u)$. Thus $\overline{\left[u (x)-u (y) \right]^{2}}\sim \Delta
(0)|x-y|^{4-d}$, which should be compared to the thermal average
$\left< \left[u (x)-u (y) \right]^{2} \right>\sim T
|x-y|^{2-d}$. Since both theories are Gaussian, the only difference is
an appearant shift in the dimension of the system from $d$ to
$d-2$. This is usually referred to as dimensional reduction.

For more than $n=1$ replicas, the theory is richer, and we will
recover the renormalization of $\Delta (u)$ itself. To this purpose,
write 
\begin{eqnarray}\label{s1}\fl
{\cal S}[u_{a},\tilde u_{a},\bar\psi_{a},\psi_{a},j_{a}]
&=&\sum_{a}  \int_{x} \Big[ \tilde u_{a}
(x) (-\nabla^{2 }u_{a} (x) +j_{a} (x)) + \bar \psi_{a} (x)
(-\nabla^{2})\psi_{a} (x) \nonumber \\
&& \qquad \qquad - \half \tilde u_{a} (x)\Delta (0)\tilde
u_{a} (x)\Big]\nonumber 
\\
&&\hspace{-0cm}-\sum_{a\neq b}\int_{x}\Big[\half\tilde u_{a} (x)
\Delta (u_{a} (x)-u_{b}
(x))\tilde u_{b} (x)\nonumber \\
&& \qquad \qquad + \half \bar \psi_{a} (x)\psi _{a} (x)\Delta''
(u_{a} (x)-u_{b} (x))\bar \psi_{b} (x)\psi_{b} (x) \nonumber \\
&&\qquad\qquad - \tilde u_{a} (x)
\Delta' (u_{a} (x)-u_{b} (x)) \bar \psi_{b} (x)\psi_{b} (x)\Big]
\ . 
\end{eqnarray}
Corrections to $\Delta (u)$ are easily
constructed by remarking that the interaction term quadratic in
$\tilde u$ is almost identical to the treatment of the dynamics in the
static limit (i.e.\ after integration over times)
\begin{equation}\label{deltaDeltafromSusy}
\diagram{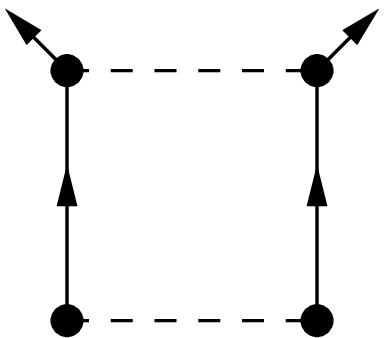}+\diagram{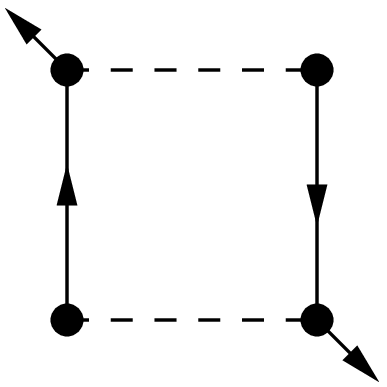} +\diagram{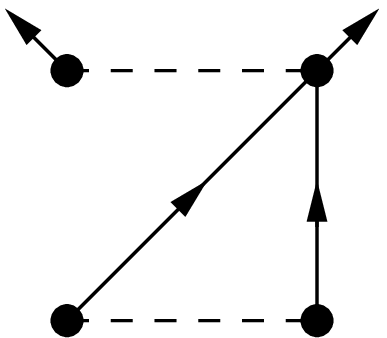} +2 \diagram{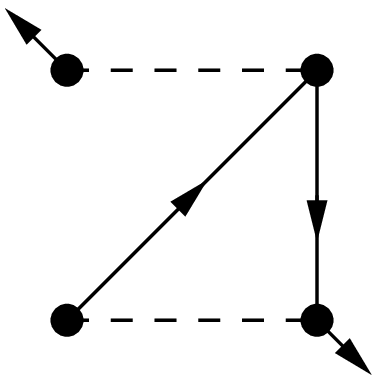}\ ,
\end{equation}
where an arrow indicates the correlation-function,
${_{x}}-\!\!\!\!\rightarrow\!\!\!\!-\!\!\!\!-\!\!\!\!-{_{y}}\, =
\left< \tilde u (x)u (y) \right>=C (x-y)$. This leads to (in the order given
above)
\begin{equation}\label{s2}
\delta \Delta (u) = \left[ - \Delta (u) \Delta'' (u)  - \Delta' (u)^{2} +
\Delta'' (u)\Delta (0) \right] \int_{x-y}C (x-y)^{2}
\end{equation}
where the last term (being odd in $u$) vanishes. Note that this
reproduces the non-linear terms in (\ref{RG1loop}). 

A non-trivial ingredient is the cancellation of the acausal loop in
the dynamics (the ``sloop'', or 3-replica term in the replica
formulation) \cite{LeDoussalWieseChauve2003}. This is provided by
taking two terms proportional to $\tilde u_{a}\Delta'
(u_{a}-u_{b})\bar \psi_{b}\psi_{b}$, and contracting all fermions:
\begin{equation}\label{sloop}
 \diagram{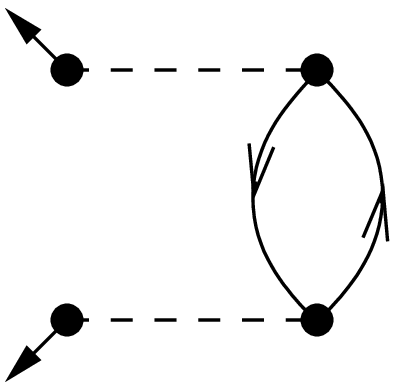} + \diagram{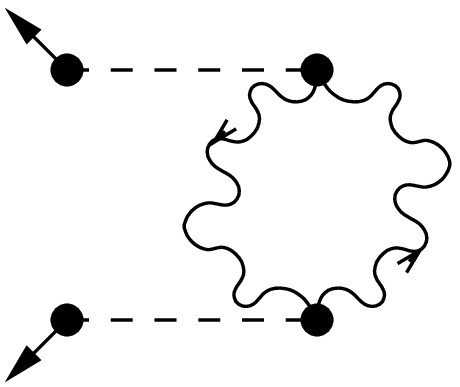} =0\ ,
\end{equation}
since the fermionic loop (oriented wiggly line in the second diagram)
contributes a factor of $-1$.

One can treat the interacting theory completely in a superspace
formulation. The action is
\begin{eqnarray}\label{super-space action}
{\cal S}[U_{a}]&=& \sum_{a}\int_{\Theta, \bar \Theta }\int_{x}
U_{a} (x,\bar \Theta ,\Theta) (\Delta_{s}) U_{a} (x,\bar \Theta
,\Theta) \nonumber \\
&&  -\half \sum_{a\neq b} \int_{x}\int_{\bar \Theta
,\Theta} \int_{
 \bar \Theta', \Theta'} R (U_{a} (x,\bar \Theta ,\Theta)-U_{b}
(x,\bar \Theta ',\Theta' ))
\ .
\end{eqnarray}
Thus non-locality in replica-space or in time is replaced by
non-locality in superspace, or more precisely in its anticommuting
component. Corrections to $R (u)$ all stem from ``superdiagrams'',
which result into bilobal interactions in superspace, not trilocal, or
higher. The latter find their equivalent in 3-local terms in
replica-space in the replica-formulation, and 3-local terms in time,
in the dynamic formulation. 

Supersymmetry is broken, once $\Delta (0)$ changes. However, a new,
shall we call it ``effective supersymmetry'', or ``scale-dependent
supersymmetry'' appears, in which the parameter $\Delta (0)$, which
appears in the Susy-transformation, changes with scale, according to
equation (\ref{flowR''0}).

An interesting question is, whether anomalous terms, proportional to
$R''' (0^{+})$ can be recovered from the supersymmetric
formulation. We show now, that  this can indeed be done, in a very
elegant way. The trick is to shift the disorder $V (u)$ which appears
in (\ref{su5}) for the bosonic part, by a small amount $\delta$ for the
fermionic part, and to take the limit of $\delta \to 0$ at the end. 
This modifies (\ref{su6b}) to
\begin{eqnarray}
%\begin{subequation}
\fl\label{su7b} &&\hspace{-.6cm}{\cal S}[u_a,j_{a},\tilde u_{a},\bar
\psi_{a},\psi_{a}]\nonumber \\
\fl&=& \sum_{a} \int_{x} \tilde u_{a} (x) (-\nabla^{2} u_{a} (x) +j_{a}
(x)) + \bar \psi_{a} (x)
(-\nabla^{2})\psi_{a} (x) \nonumber \\
\fl&&+ \sum_{a,b}\int_{x} \Big[ \half \tilde u_{a} (x)\Delta (u_{a} (x)-u_{b}
(x))\tilde u_{b} (x)+ \half \bar \psi_{a} (x)\psi _{a} (x)\Delta''
(u_{a} (x)-u_{b} (x))\bar \psi_{b} (x)\psi_{b} (x) \nonumber \\
\fl&&\qquad\qquad - \tilde u_{a} (x)
\Delta' (\delta +u_{a} (x)-u_{b} (x)) \bar \psi_{b} (x)\psi_{b}
(x)\Big]\\
\label{s3}
\fl&=&\sum_{a} \int_{x} \Big[ \tilde u_{a} (x) (-\nabla^{2} u_{a} (x) +j_{a}
(x)) + \bar \psi_{a} (x)
(-\nabla^{2})\psi_{a} (x) 
+\half \tilde u_{a} (x)^{2} \Delta (0) \nonumber \\
\fl&& \qquad\qquad - \tilde u_{a} (x) \Delta'
(\delta) \bar \psi_{a} (x)\psi_{a} (x) \Big]
\nonumber \\
\fl&&+ \sum_{a\neq b}\int_{x} \Big[ \half \tilde u_{a} (x)\Delta (u_{a}
(x)-u_{b} (x))\tilde u_{b} (x)+ \half \bar \psi_{a} (x)\psi _{a}
(x)\Delta''
(u_{a} (x)-u_{b} (x))\bar \psi_{b} (x)\psi_{b} (x) \nonumber \\
\fl&&\qquad\qquad - \tilde u_{a} (x)
\Delta' (\delta +u_{a} (x)-u_{b} (x)) \bar \psi_{b} (x)\psi_{b}
(x)\Big] 
\ .
%\end{subequation}
\end{eqnarray}
Now the term with $a=b$ is also well defined; the last term of
(\ref{su7b}), as made explicit in (\ref{s3}), is of the form:
\begin{equation}\label{su8} \fl
\int_{x}\sum_{a} -\tilde u_{a} (x) \Delta' (\delta) \bar \psi_{a}
(x)\psi_{a} (x) - \sum_{a\neq b} \tilde u_{a} (x) \Delta' (\delta
+u_{a} (x)-u_{b} (x)) \bar \psi_{b} (x)\psi_{b} (x)\ .
\end{equation}
To demonstrate how this can be put into use, let us calculate the
contribution to the 2-point function at 1-loop order, which is
naively ambiguous \cite{ChauveLeDoussalWiese2000a,LeDoussalWieseChauve2002,LeDoussalWieseChauve2003}:
\begin{eqnarray}\label{su9}
\fl\delta_{\mathrm{1loop}}\left< u_{a} (q) u_{a} (-q) \right>&=&
\diagram{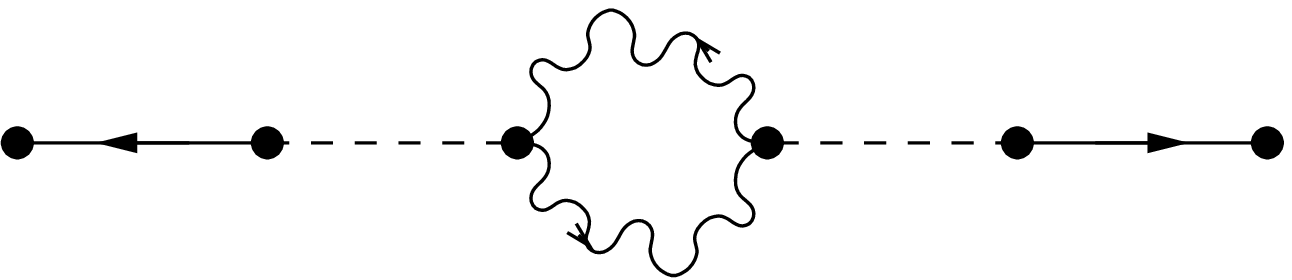}\nonumber \\
&=& - \frac{1}{(q^{2}+m^{2})^{2}}\, \Delta'
(\delta)^{2}\int_{p} \frac{1}{(p+q/2)^{2}+m^{2}}
\frac{1}{(p-q/2)^{2}+m^{2}}
\ .
\end{eqnarray}
Note that the minus-sign comes from the closed fermion loop. In the
limit of $\delta \to  0$, this gives
\begin{equation}\label{su10}\fl
\delta_{\mathrm{1loop}}\left< u_{a} (q) u_{a} (-q) \right>= -
\frac{1}{(q^{2}+m^{2})^{2}} \Delta' (0^{+})^{2}\int_{p}
\frac{1}{(p+q/2)^{2}+m^{2}} \frac{1}{(p-q/2)^{2}+m^{2}} \ .
\end{equation}
Also note that this would equivalently work for a $N$-component field
$\vec u=\left\{u_{a} \right\},$ $a=1\dots N$, a case which poses
additional difficulties, since derivatives have to be taken in a given
direction \cite{BalentsDSFisher1993,LeDoussalWiesePREPd}.

Even higher correlation functions are immediate. E.g.~is 
\begin{equation}\label{su11}
\Gamma_{\tilde u_{a}\tilde u_{a}\tilde u_{a}\tilde u_{a}}
=\diagram{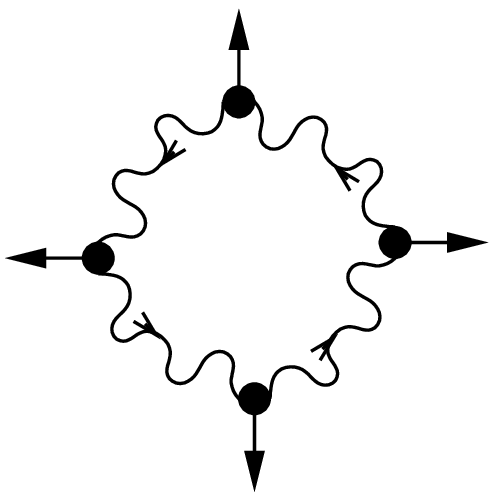}= - \frac{\Delta' (0^{+})^{4}}{64} \int_{q}\frac{1}{(
q^{2}+m^{2})^{4}} \ .
\end{equation}
This has first been obtained, using straightforward dynamical
perturbation theory in \cite{LeDoussalWiese2003a}. While the principle
is simple, the calculations are actually very cumbersome. The
sloop-method \cite{LeDoussalWieseChauve2003} is another efficient
approach. (See also   \cite{BalentsLeDoussal2004}.)

Let us finally mention that the method correctly calculates the
anomalous term of the ``Mercedes-star'' diagram at 3-loop order
\cite{LeDoussalWiesePREPb}, 
\begin{equation}\label{MercedesStar}
\delta R (u) = \half \left[R''' (u)^{4} -R'''(u)^{2} R'''
(0^{+})^{2}\right] \diagram{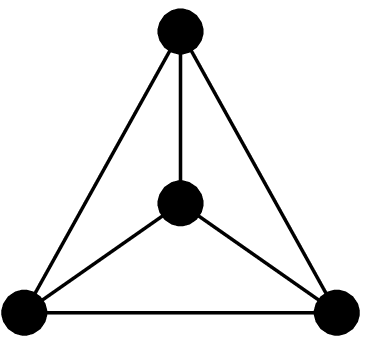}\ ,
\end{equation}
where the icon stands for the  momentum-integral only. 
The correction to $\Delta (u)=-R'' (u)$, and picking the term proportional to
$\Delta''' (u) =-R^{(5)} (u)$ comes from 
\begin{equation}\label{s4}
\delta \Delta (u)= \dots + \diagram{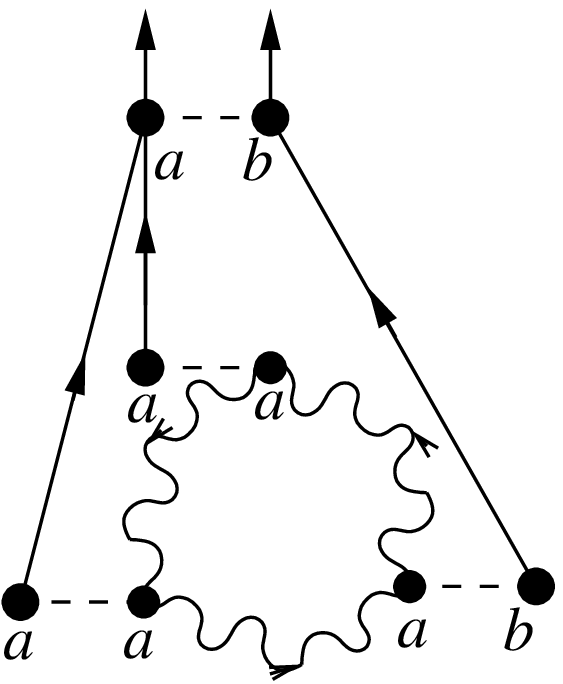} =\left[\dots + \Delta'
(0^{+})^{2} \Delta' (u)\Delta''' (u) \right]\diagram{3loopi}
\end{equation}
Note that two of the vertices are at argument $\Delta' (0^{+})$.  This
can otherwise only be calculated within the sloop-method.

\section{Relation between Supersymmetry Breaking, Functional RG,  and
Replica-Symmetry Breaking}\label{s:FRG-Susy-RSB}
Another popular approach to disordered systems is the
replica-variational method, invoking replica-sym\-metry breaking (RSB). 
This method has for the problem at hand been developed in
\cite{MezardParisi1991}. It consists in making the replacement 
\begin{equation}
\sum_{a,b}
\tilde u_a (x)\Delta (u_{a} (x)-u_{b} (x)) \tilde u_{b} (x) \
\longrightarrow \ \sum_{a,b} \tilde u_{a} (x) \tilde u_{b} (x) \sigma_{ab} \ . 
\end{equation}
This approximation is valid in the limit of a field $u$ with an
infinite number $N$ of components; so for the following discussion we
have to restrain ourselves to that limit. The variational replica
approach then makes an ansatz for $\sigma_{ab}$, with different
correlations $\sigma_{ab}$ between different pairs of
replicas. Finally a variational scheme is used to find an optimal
$\sigma_{ab}$. The result (in the case of long-range correlated
disorder, where the comparison can be made \cite{LeDoussalWiese2003b})
is a hierarchic matrix with an infinite number of different
parameters, of the form
\begin{equation}\label{RSBsigma}
\sigma = \left(\parbox{3cm}{\fig{3cm}{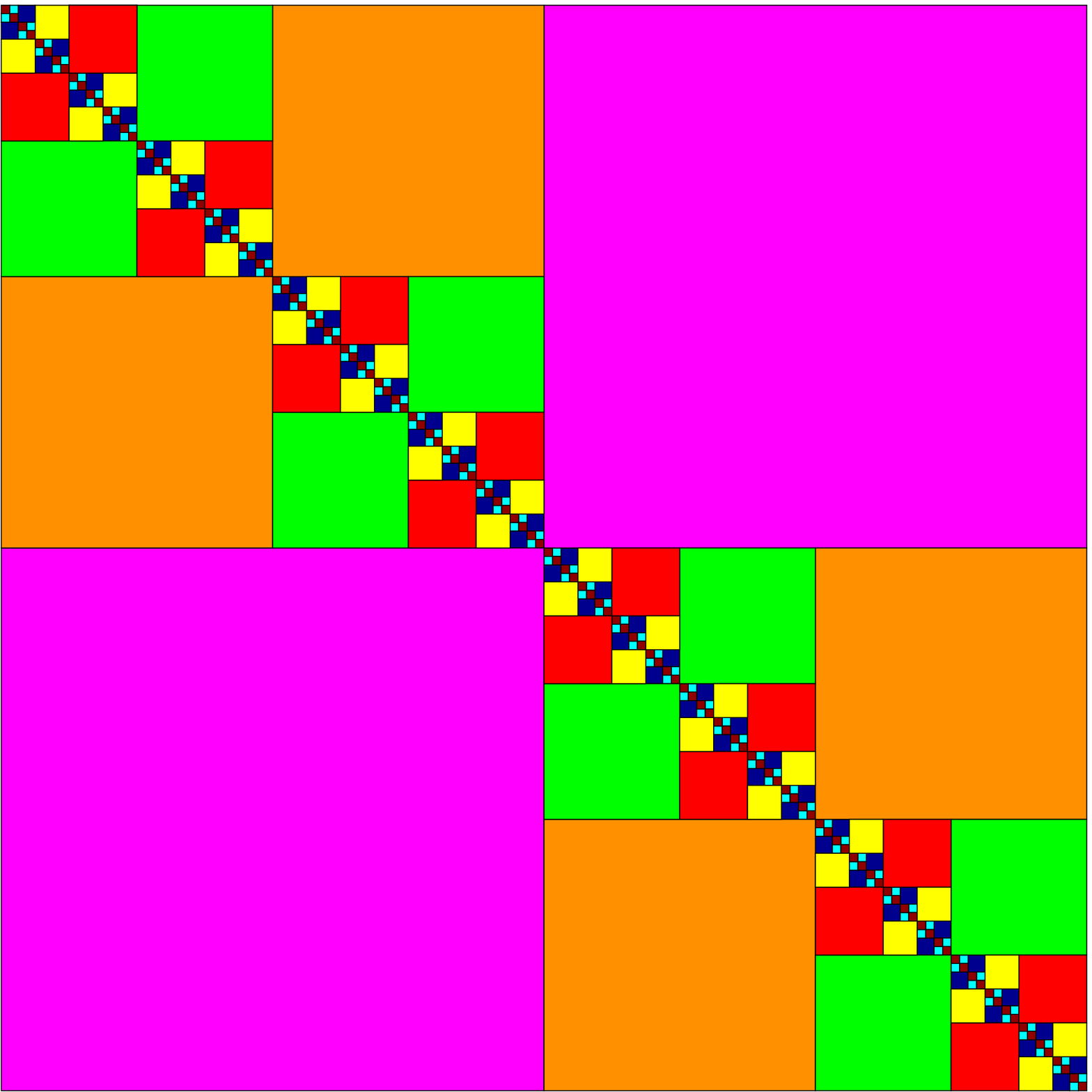}} \right)\ .
\end{equation} 
The exact form, and how it can be parameterized by a continuous
function $[\sigma] (z)$, $0\le z\le 1$ is not of importance for the
following.

What is important is that RSB appears exactly at the same moment as in
the functional RG the cusp appears
\cite{BalentsBouchaudMezard1996,LeDoussalWiese2001,LeDoussalWiese2003b}.
Moreover, there is a precise relation between the 2-point function
calculated by the RSB and FRG methods
\cite{LeDoussalWiese2001,LeDoussalWiese2003b}. The latter do not
coincide, since the calculations are implicitly done in zero external
field (RSB) and vanishing external field (FRG), leading to physically
different situations (as for a standard ferro-magnet).

In section \ref{a:susy}
we have shown, that this is also the moment, when supersymmetry is
broken. While the treatment there was for a 1-component field, the
conclusion is the same for an $N$-component field, and persists in the
limit of $N\to \infty$. We can thus conclude that the breaking of
supersymmetry, of replica symmetry, and the appearance of the cusp
are all but different manifestations of the same underlying physical
principle: the appearance of multiple minima.

\section{Acknowledgments} It is a pleasure to thank Andreas Ludwig 
for stimulating discussions on the supersymmetric method.

The work on the functional RG is part of  a series of
inspiring ongoing collaborations with Pierre Le Doussal, for which I
would like to express my gratitude. 

I am most grateful to Lothar Sch\"afer for years of continuous
support, for always being available to discuss and for persisting
until each problem had  been clarified. His enthusiasm even at
the oddest hours were an enormous encouragement.  I will always be in
his debt.

%\bibliography{../../citation/citation}

\end{document}